\documentclass[%
reprint,
preprintnumbers,
nofootinbib,
 amsmath,amssymb,
 aps,
]{revtex4-2}

\usepackage{graphicx}
\usepackage{dcolumn}
\usepackage{tabularx}
\usepackage{bm}
\usepackage{color}

\begin{document}

\title{A simple non-parametric reconstruction of parton\\distributions from limited Fourier information}

\author{Herv\'e Dutrieux}
 \email{hldutrieux@wm.edu}
\affiliation{Physics Department, William \& Mary, Williamsburg, VA 23187, USA
}
\author{Joseph Karpie}
 \email{jkarpie@jlab.org}
\affiliation{Thomas Jefferson National Accelerator Facility, Newport News, VA 23606, USA 
}
\author{Kostas Orginos}
 \email{kostas@jlab.org}
\affiliation{Physics Department, William \& Mary, Williamsburg, VA 23187, USA \\ Thomas Jefferson National Accelerator Facility, Newport News, VA 23606, USA
}
\author{Savvas Zafeiropoulos}
 \email{savvas.zafeiropoulos@cpt.univ-mrs.fr }
\affiliation{Aix Marseille Univ, Universit\'e de Toulon, CNRS, CPT, Marseille, France.
}

\date{\today}

\begin{abstract}
Some calculations of parton distributions from first principles only give access to a limited range of Fourier modes of the function to reconstruct. We present a physically motivated procedure to regularize the inverse integral problem using a Gaussian process as a Bayesian prior. We propose to fix the hyperparameters of the prior in a meaningful physical fashion, offering a simple implementation, great numerical efficiency, and allowing us to understand and keep control easily of the uncertainty of the reconstruction.
\end{abstract}

\preprint{JLAB-THY-24-4242}
\maketitle

\section{Introduction}

Parton distributions are fundamental quantities in our understanding of hadronic structure, one of the most active topics of particle physics for the past 50 years. They are notoriously difficult to compute because of the strongly coupled interaction of quarks and gluons at low energy. In recent years, they have been thoroughly studied in lattice quantum chromodynamics (QCD) from non-local matrix elements of the type \cite{Ji:2013dva, Ji:2014gla, Radyushkin:2017cyf, Orginos:2017kos}:
\begin{align}
&\langle P | \bar{\psi}\left(-\frac{z}{2}\right)\gamma^\mu \hat{W}\left[-\frac{z}{2},\frac{z}{2};A\right]\psi\left(\frac{z}{2}\right) | P \rangle \nonumber \\
&\hspace{40pt}= P^\mu M(\nu,z^2)+z^\mu N(\nu,z^2)\,, \label{eq:matelem}
\end{align}
where $z$ is a space-like separation, $P$ the hadron momentum and $\nu = P \cdot z$ a Lorentz invariant often called Ioffe time. Assuming a proper removal of UV divergences, $M(\nu)$ can be related in the short-distance factorization scheme to the parton distribution $f_q(x)$ through the Fourier transform:
\begin{equation}
M(\nu) = \int_{-1}^1 \mathrm{d}x\,e^{ix\nu}f_q(x)\,.
\end{equation}
We ignore the matching procedure which bears no impact on the inverse problem we are concerned with. In state-of-the-art calculations, the range in $\nu$ accessible with controlled higher twist uncertainties barely reaches 10. See \cite{Constantinou:2020hdm,Constantinou:2020pek,Lin:2023kxn} for reviews on the state of lattice computations. The full reconstruction of $f_q(x)$ therefore suffers from an ill-posed inverse problem. It is typically regularized by parametrizing $f_q(x)$ with a few free parameters -- at the risk of excessive model dependence -- or a neural network \cite{Cichy:2019ebf,Karpie:2019eiq,DelDebbio:2020rgv,Gao:2022iex,Khan:2022vot} whose behavior may be difficult to explain in a physically motivated way. Other methods, such as the Backus-Gilbert algorithm \cite{BG, Hansen:2019idp}, maximum entropy method \cite{Asakawa:2000tr}, Bayesian reconstruction \cite{Burnier:2013nla}
or polynomial expansions \cite{Karpie:2021pap} have been studied. Early comparisons of various $x$-dependent reconstruction methods can be found in \cite{Karpie:2019eiq, Liang:2019frk}. 

We will use in this paper the formalism of Gaussian processes, already explored in the context of lattice parton distributions in \cite{Alexandrou:2020tqq}. We follow the idea of a non-parametric reconstruction using a Gaussian process as Bayesian prior in $x$-space advocated in \cite{Candido:2024hjt}, extending on the previous works \cite{DelDebbio:2021whr,Candido:2023nnb}. Gaussian processes have already been used in lattice QCD computations for other inverse problems, such as in \cite{Horak:2021syv, Horak:2023xfb}, as well as in non-perturbative studies in continuum QCD \cite{Pawlowski:2022zhh}.  They offer many advantages in terms of flexibility and proper mathematical treatment of the uncertainty. However, as with any regularization of an inverse problem, this technique introduces a bias, here in the form of a choice of kernel and hyperparameters. In \cite{Candido:2024hjt} it is proposed to infer the hyperparameters from the fitted dataset, however, we find it advantageous for the specific inverse problem at hand to propose a physically motivated variant of this idea where the hyperparameters are entirely fixed by easily tuned physical constraints.  Especially when the range in $\nu$ is small and the available fitted information is very limited, we believe that it is useful and more meaningful to keep direct physical control over the hyperparameters and therefore the uncertainty. In a subsequent publication, we will study the effects of hyperparameter optimizations and compare these approaches with the simple and intuitive method proposed here. The approach presented here is also highly efficient from a numerical point of view and has the advantage of preserving the normal distribution of the reconstruction. Therefore, the results of the reconstruction can be shared in an efficient and exact fashion. Although we present our results in the context of the short-distance factorization of lattice matrix elements, many elements of the discussion presented here apply to a variety of other inverse problems both in lattice and phenomenological determinations of parton distributions.

The article is organized in the following way. We first remind the principles of using a Gaussian process as Bayesian prior to regularize a functional inverse problem. Then we present our procedure and propose a closure test, detailing notably the effect of the range in $\nu$ on the uncertainty. We then apply the method to actual lattice QCD data, comparing its results to a usual parametric fit. We end up with a discussion of how other important inverse problems of parton distributions in lattice QCD and phenomenology may or may not benefit straightforwardly from our physically motivated procedure.

\section{Gaussian processes as Bayesian priors}

Let us assume that we possess experimental information on the function $f_q$ through some linear operators parametrized by $\nu$:
\begin{equation}
M(\nu) = B(\nu) \circ f_q\,.
\end{equation}
Our canonical example will be that $M(\nu)$ is a set of discrete Fourier modes of $f_q$ and $B(\nu)$ the linear operator $\int \mathrm{d}x\,e^{ix\nu}$. Other examples of linear inverse problems relevant for our physics are evoked in the final section of the document. If $f_q$ is just any function and the set $(\nu)$ is finite, there is obviously an infinite number of possible reconstructions. A large number of regularization techniques rely on Bayesian priors. One assumes prior information $I$ on the credibility of various models, represented by the prior probability distribution $\mathbb{P}(f_q | I)$. Then Bayes' theorem gives,
\begin{equation}
\mathbb{P}(f_q | M, I) = \frac{\mathbb{P}(M | f_q)\mathbb{P}(f_q | I)}{\mathbb{P}(M | I)}\,,
\end{equation}
where $\mathbb{P}(M | f_q)$ is the data likelihood, $\mathbb{P}(M | I)$ is the evidence, and $\mathbb{P}(f_q | M, I)$ is the posterior  knowledge of $f_q$ given the experimental data $M$ and the prior information $I$.
  For instance, the expectation value $\langle f_q(x) \rangle$ is obtained by:
\begin{equation}
\langle f_q(x) \rangle = \int \mathrm{d}f_q\, f_q(x) \mathbb{P}(f_q | M,I)\,.
\end{equation}

A Gaussian process \cite{Ras} is an infinite collection of random variables indexed by a continuous parameter, such that any finite subset follows a multivariate normal distribution. Gaussian processes are fully characterized by their mean $g(x)$ and covariance function $K(x, x')$. They are the natural extension of the normal distribution to continuous degrees of freedom, and form therefore an ideal Bayesian prior for a functional problem:
\begin{align}
&\mathbb{P}(f_q|I) = (\det(2\pi K))^{-1/2} \exp\bigg(-\frac{1}{2}\times \nonumber \\&\hspace{10pt}\int \mathrm{d}x \mathrm{d}x' \ [f_q(x) - g(x)] K^{-1}(x, x') [f_q(x') - g(x')]\bigg)\,,
\end{align}
where $K^{-1}$ is the inverse covariance function. The rigorous definition of the functional determinant $\det(2\pi K)$ requires an extension of the notion of matrix determinant, performed either with the zeta operator or a discretized path integral formulation \cite{OSGOOD1988148}. For all practical purposes, since numerical computations require a finite number of degrees of freedom, we will evaluate the Gaussian process on a grid of points $(x_i)$. It then becomes an ordinary multivariate normal distribution. The covariance function $K(x, x')$ is replaced by a covariance matrix $K_{ij} = K(x_i, x_j)$ whose determinant can be evaluated without issues, and the linear operator $B(\nu)$ becomes a matrix $B_{ki}$ that maps $(x_i)$ into $(\nu_k)$. The values of the posterior outside of the grid are obtained by a linear interpolation from the $f_q(x_i)$. If one considers the traditional Gaussian likelihood of the data given the model $\mathbb{P}(M | f_q)$, then the posterior distribution on the $(x_i)$ grid is also a multivariate normal distribution that can be computed analytically. 

With these conventions, the posterior mean is \cite{DelDebbio:2021whr, Candido:2024hjt}:
\begin{equation}
\langle f_q(x_i) \rangle = g(x_i) + KB^T [C + BKB^T]^{-1} [M - B g] \label{eq:postmean}
\end{equation}
where $C$ is the covariance of the data  $M$, and the covariance matrix of the posterior is:
\begin{align}
H &= (K^{-1} + B^T C^{-1} B)^{-1}\,,\\
&= K - KB^T[C+BKB^T]^{-1} BK\,.
\label{eq:postcov}
\end{align}

\section{The method}

First, we start by separating the even and odd parts of the $x$-dependence (which effectively corresponds to selecting different combinations of quark and anti-quark distributions), so that our functions can be modeled only in the $x \in [0, 1]$ interval. Since we are using a Gaussian process as Bayesian prior, we then merely need to specify a mean $g(x)$ and a covariance function $K(x, x')$. Many proposals are available in the literature. A successful parametric modeling of parton distributions is produced by variants of:
\begin{equation}
x^\alpha (1-x)^\beta \times \textrm{ smooth function of }x\,,
\end{equation}
where $\alpha$ is typically slightly negative, producing a power divergence when $x \rightarrow 0$, and $\beta = 2 \sim 3$. See \cite{Ball:2016spl} for a critical review of this parametric model in the context of global fits, where the authors warn against potentially unrealistic uncertainty in extrapolation regions. In our inverse problem, especially when the range in $\nu$ is small, this parametric form introduces a considerable bias in the reconstruction at endpoints. Sensitivity to point-like features, such as the exponent of the behavior at $x = 0$ or $x = 1$, is only achieved in the limit of large Fourier frequencies. The low frequencies only characterize low-resolution information on the function. To fulfill our premise of non-parametric reconstruction, we will abstain from making any reference to that model in our reconstruction.

Instead, we choose a covariance function based only on general physical considerations:
\begin{equation}
K(x, x') = \sigma^2 \exp\left(-\frac{(\ln(x) - \ln(x'))^2}{2 l^2}\right)\,, \label{eq:kernellog}
\end{equation}
where $\sigma^2$ controls the variance of our prior and $l$ is a logarithmic correlation length in $x$ space which enforces the smoothness of the reconstruction. By using $\ln(x)$ in the numerator, we have introduced the prior expectation that the physics at $x \ll 1$ and $x$ close to 1 are fundamentally different and should be decorrelated. This also enforces a requirement of smoothness of the reconstruction if the ratio $x / x'$ lies typically within $[e^{-l}, e^l]$. This choice of kernel is therefore imprinting the first physics-driven regularization of our inverse problem.

But how should we choose the values of the prior's hyperparameters $(\sigma^2, l)$ and mean $g(x)$ which we have yet to decide upon? A proposal explored in \cite{Candido:2024hjt} is to sample the hyperparameters according to the evidence $\mathbb{P}(M|I)$. In practice, it means that \textit{e.g.} the correlation length and variance of the prior in $x$-space would be inferred from the dataset itself. However, in the case of our inverse problem, we have no data constraint in $x$-space, only low-order Fourier frequencies. It is not obvious why information of smoothness derived from the low-order Fourier frequencies provides a good guide to reconstructing the higher frequencies. From a theoretical point of view, it has been long argued \cite{Braun:1994jq} that different Ioffe times $\nu$ correspond to a different physics of hadronic structure. Empirically, we know (and will verify again in this document) that the low-order frequencies constrain well the reconstruction at intermediate values of $x$, and the high-frequencies mostly constrain the behavior in the limit $x \rightarrow 0$. Therefore, using a smoothness information obtained at low frequency to constrain the high frequency may effectively go against our goal of decorrelating small and large $x$ regions of the reconstructed function. Various empirical tests of hyperparameter sampling lead, especially when the range in $\nu$ is small, to larger prior correlation lengths and smaller prior variances than physical insights would lead us to, as will be elaborated in more detail in a subsequent publication. Already in the study presented in \cite{Candido:2024hjt} with synthetic lattice data spanning in the range $\nu \in [0, 8]$, it appears that the correlation length is very poorly informed by the data and is therefore free to err to as large values as allowed by the prior put on the hyperparameter. Direct comparison is however made difficult by the use of a different kernel in that study. Such features may give the wrong intuition that a reduced range in $\nu$ produces overall good $x$ reconstruction.

Finally, the procedure demonstrated in this work is extremely numerically efficient compared for instance to sampling the posterior of GP hyperparameters, or even the traditional non-linear minimization of parameterized functions. By fixing the hyperparameters, the posterior is simply a Gaussian whose mean and covariance are analytically derived through eqs. \eqref{eq:postmean}-\eqref{eq:postcov}. This setting is a specific case of sampling hyperparameters with delta functions for the hyperparameter priors. The advantage of loosening those hyperparameter priors is to relax some of the choices made by the scientist, at the increased computational cost of Markov Chain Monte Carlo sampling the non-trivial posterior. Even so, some choice of hyperparameter priors must be made and those biases can never be completely removed. In this work, we propose a simple set of data-driven choices for analyzing the Ioffe time distributions which will give efficient results without requiring sampling. If one wishes to augment the dataset with other observables such as those discussed in Sec.~\ref{sec:other_inverse}, a single good set of physical data-driven choices may not be clear and sampling of the parameters may be necessary.

\paragraph*{}
\paragraph*{\textbf{Procedure}}
We suggest:
\begin{enumerate}
\item Set the logarithmic correlation length to $l = \ln(2)$. 

\item Always keep $g(x) = \sigma$.

\item Fix $\sigma^2$ such that the maximal standard deviation of the Fourier transform of the posterior for any $\nu \in [\nu_{max}, +\infty]$ is exactly $\textrm{max}(1.3 \Delta M(\nu_{max}), 0.3 |M(\nu_{max})|)$, where $\nu_{max}$ is the maximal value of $\nu$ in the dataset\footnote{Since the parton distributions we are concerned with are absolutely integrable, their Fourier transform tends towards 0 as $\nu \rightarrow \infty$. Therefore, in practice, the criterion can be checked on a fairly narrow range in $\nu$. For the application to real data, $[\nu_{max}, 50]$ proved plenty enough. For the closure test with extended range in Ioffe time, $[\nu_{max}, 150]$ was enough.}. 
\end{enumerate}

\begin{figure}
\centering
\includegraphics[width=0.75\linewidth]{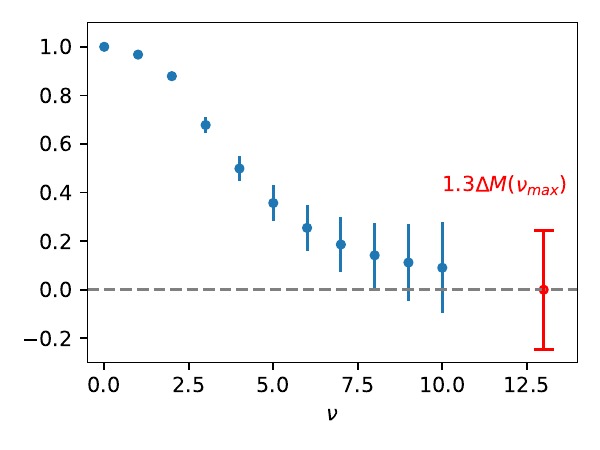}
\includegraphics[width=0.75\linewidth]{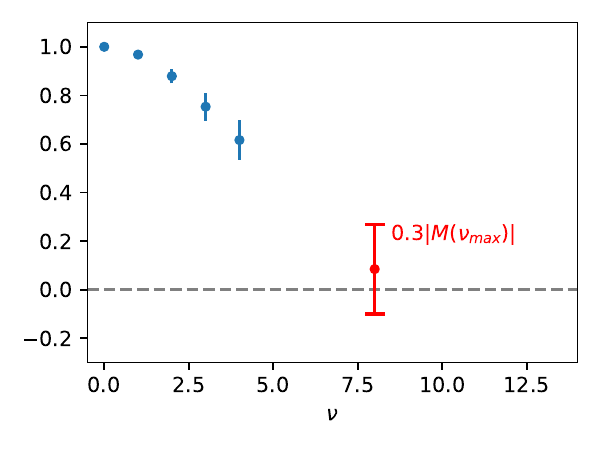}
\caption{Explanation of our targeted reconstruction uncertainty in the extrapolation region 1) increase of the last uncertainty by 30\% and 2) when there is a clear non-vanishing signal in the extrapolation region.}
    \label{fig:extrapol_unc}
\end{figure}

\paragraph*{\textbf{Explanations}}
\begin{enumerate}
\item A logarithmic correlation length of $\ln(2)$ corresponds to starting to decorrelate $x = 1$ from $x = 0.5$, $x = 0.5$ from $x = 0.25$, etc. This is a physics-driven prescription on the degree of flexibility that one expects in ordinary parton distributions. The dataset itself may have features that enforce a stronger smoothness than what this baseline correlation length requires, through the data likelihood term in addition to the effects of the prior.

\item By picking $g(x) = \sigma$, we introduce by definition a prior which is uniformly one standard deviation above 0 for all $x$. Because our data is not informative on the behavior at the point $x = 0$ and the prior covariance kernel enforces no correlation in this limit, the posterior is exactly equal to the prior when $x = 0$. Therefore, the posterior will be just one standard deviation above 0 at $x = 0$. This allows us to enforce a loose positivity prior on the reconstruction and to reflect the large uncertainty at small $x$, which is fixed by design at 100\% when $x = 0$.
\item We want the uncertainty to increase in the extrapolation region away from $\nu_{max}$. One has typically no control over this feature when sampling over $(\sigma^2, l)$ according to the evidence $\mathbb{P}(M | I)$. Yet, a visual inspection of the uncertainty in the extrapolation region is a powerful empirical criterion to evaluate whether an $x$-space reconstruction is too optimistic or pessimistic in its uncertainty quantification. Deciding on what is a reasonable degree of increase in uncertainty in an extrapolation does not admit a simple good answer. 
We have chosen to require the maximum uncertainty in the extrapolation region $[\nu_{max}, +\infty]$ to be $30\%$ larger than the uncertainty of the point $M(\nu_{max})$.
This gives the constraint $1.3 \Delta M(\nu_{max})$. Since the uncertainty tends to 0 when $\nu \rightarrow \infty$, a point in the extrapolation region with maximal uncertainty always exists, and it is reached for practical purposes fairly quickly in $\nu$. To satisfy the criterion, we suggest starting from a small value of $\sigma$ and increasing it gradually until the criterion is satisfied. Since one reconstruction takes typically of the order of one second, a correct result is found extremely easily.   A visual depiction of this targeted uncertainty in the extrapolation region is given in the upper plot of Fig. \ref{fig:extrapol_unc}. However, the dataset may end while there is still a non-vanishing signal in the Fourier space that lies within the extrapolation region. The lower plot of Fig. \ref{fig:extrapol_unc} highlights our request in uncertainty in that case, of $0.3 |M(\nu_{max})|$.
\end{enumerate}

We summarize our physical regularization in Table~\ref{tab:tab_motiv}. Of course, some of the choices that we have made here may not be consensual for some readers. The most problematic is likely the choice of $\sigma^2$ which is dictated by a \emph{perception} of what a good extrapolation uncertainty is. We have found our parameters to give a satisfactory uncertainty reconstruction without requiring any tuning in the cases where we have tried them. But if one is unsatisfied by the extent of the extrapolation uncertainty, for instance, it is very easy to act on the hyperparameters accordingly. Let us stress again the important philosophy underlying this reconstruction procedure: when performing an extrapolation in a region where there is no data, and where we have a physical intuition that the physics is fairly different compared to the region where we have data, then there is no objective choice of reconstruction procedure and choices must be made.

\begin{table}[]
\centering
\bgroup
\def\arraystretch{2.5}
\begin{tabular}{c|c}
\textbf{Choice} & \textbf{Motivation}  \\
\hline
$K(x,x')$ eq. \eqref{eq:kernellog}  & \parbox{5cm}{Enforce smoothness in $x$ while decorrelating small $x$ from large $x$} \\ \hline
$l = \ln(2)\approx 0.693$ & \parbox{5cm}{Set the flexibility of the $x$-dependence} \\ \hline
$g(x) = \sigma$ & \parbox{5cm}{Set the uncertainty to 100\% at $x=0$ and loosely enforce positivity} \\ \hline
\parbox{3cm}{$\sigma^2$ such that the worst uncertainty in Fourier space is $\textrm{max}(1.3 \Delta M(\nu_{max}),$ $0.3 |M(\nu_{max})|)$} & \parbox{5cm}{Control the size of uncertainty in the extrapolation region in Fourier space}
\end{tabular}
\caption{Summary of our hyperparameters and the motivation behind our physical prescription.}
\label{tab:tab_motiv}
\egroup
\end{table}

The alternative strategy of allowing hyperparameters to be sampled in priors beyond the fixed (delta function) prescription we suggest only mildly relaxes the arbitrariness of these choices. Since some hyperparameters are very poorly controlled by the data, the choice of the hyperparameter priors may become itself a source of systematic uncertainty. 
On the other hand, the hyperparameters that are controlled by the data and affect significantly the extrapolation region (high-frequency region), are imprinting low-frequency information on the high-frequency domain, mixing up two physically different regimes. As a result, we believe that it is more difficult to control exactly the impact of the physical assumptions on the uncertainty when sampling is performed, compared to a direct fixing procedure as we propose.
Our upcoming work will make direct comparisons of these two approaches to better understand the issues involved.

Finally, let us mention that we could easily enforce the requirement that the distribution vanishes at $x = 1$. This can be performed for instance by adding a row to the $B$ matrix which only selects $f_q(1)$, and enforcing its value to be 0. In the following tests, we have chosen to let it free to see what constraining power the data has on the behavior when $x \rightarrow 1$.

\section{Closure test}

\begin{figure}
    \centering
    \includegraphics[width=0.99\linewidth]{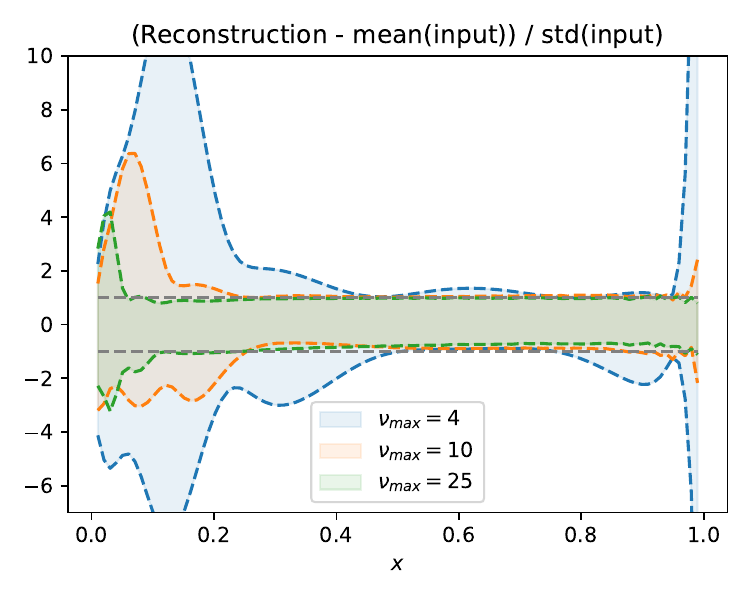}
    \caption{Comparison of the posterior reconstruction for different values of $\nu_{max}$ with respect to the input distribution. A perfect reconstruction of the input is represented by the dotted gray lines at $\pm 1$.}
    \label{fig:zdist}
\end{figure}

\begin{figure*}
\centering
\includegraphics[width=0.8\linewidth]{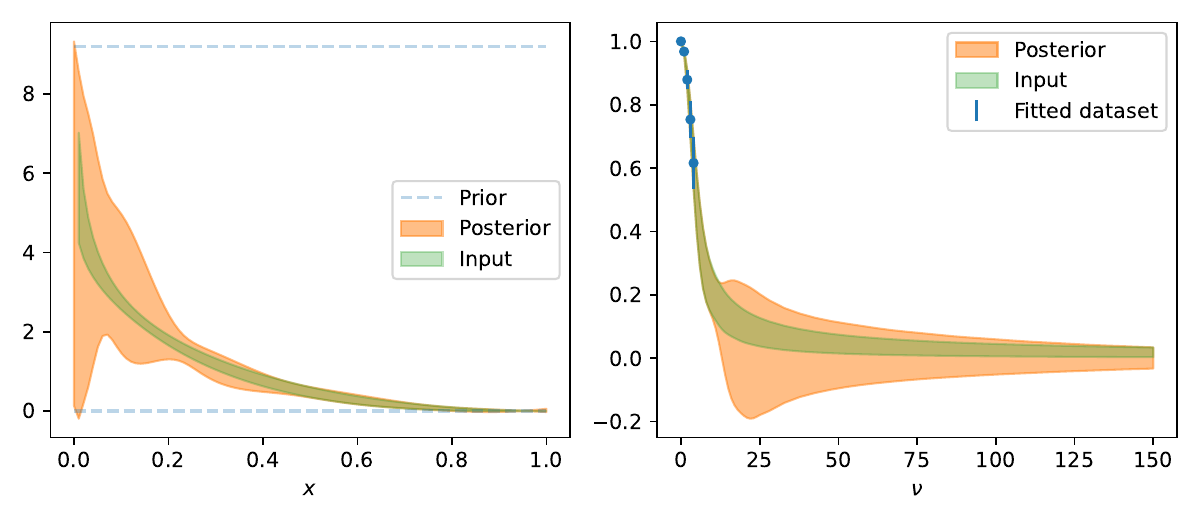}
\caption{The input (green band) represents the known distribution used to generate the fitted dataset (blue points). The posterior (orange band) is the reconstruction obtained from the fitted dataset and our prior selection. The range in $\nu$ of the fitted dataset is small, resulting in large uncertainties in the $x$ reconstruction.}
    \label{fig:closure_test1}
\end{figure*}

\begin{figure*}
\centering
\includegraphics[width=0.8\linewidth]{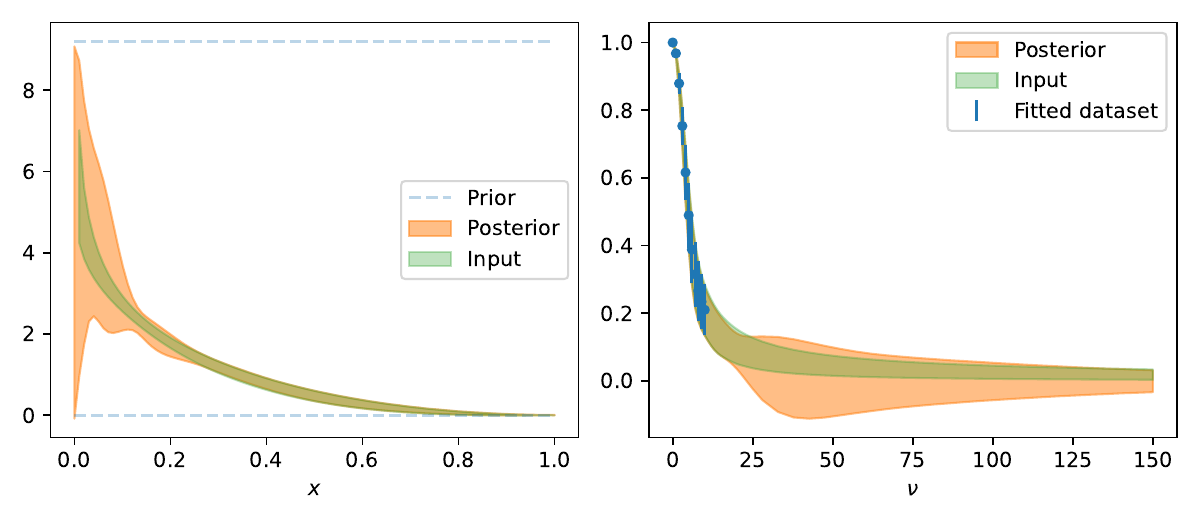}
\caption{See caption of Fig. \ref{fig:closure_test1}. This range in $\nu$ is representative of typical state-of-the-art lattice calculations.}
    \label{fig:closure_test2}
\end{figure*}

\begin{figure*}
\centering
\includegraphics[width=0.8\linewidth]{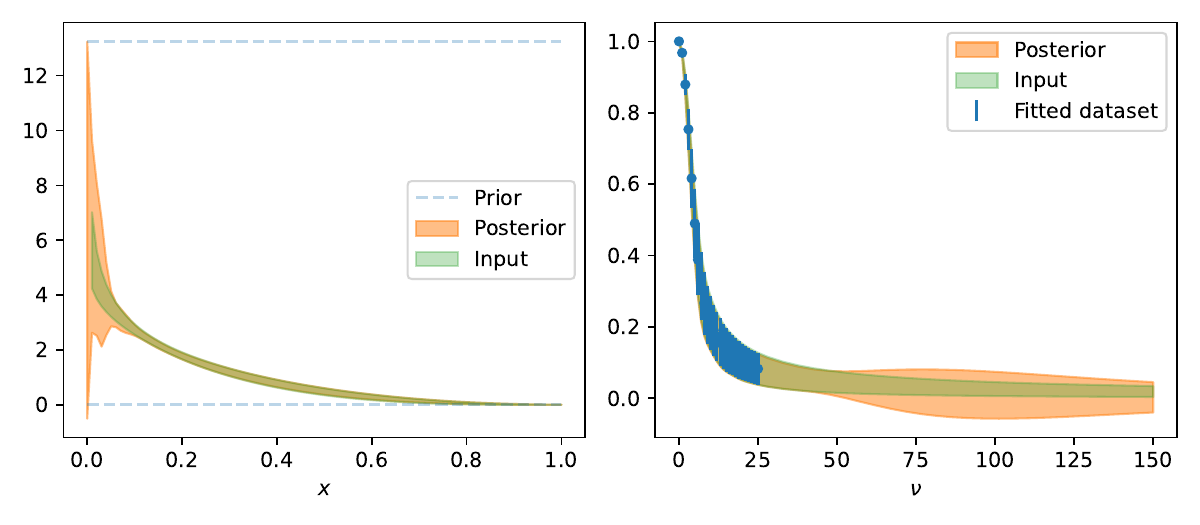}
\caption{See caption of Fig. \ref{fig:closure_test1}. The range in $\nu$ extends considerably beyond the region where higher-twist uncertainties are considered controlled in the non-local operator within current lattice calculations.}
    \label{fig:closure_test3}
\end{figure*}

To test the effect of our procedure in a controlled environment, we generate 1000 functional samples of the input distribution $f_q(x) = N x^\alpha (1-x)^\beta$ where $\alpha$ and $\beta$ follow normal distributions of respective (mean, standard deviation): $(-0.2, 0.1)$ and $(2.5, 0.5)$. $N$ is fixed by the constraint $\int_0^1 \mathrm{d}x\,f_q(x) = 1$. We produce three datasets of the real part of the Fourier modes to study the impact of the range in $\nu$ on the reconstruction: $\nu \in [0, \nu_{max}]$ with $\nu_{max} = \{4, 10, 25\}$. The grid $(x_i)$ is made of 100 points evenly spaced between $10^{-6}$ and 1. The lower bound is non-zero to allow the evaluation of the prior covariance kernel eq. \eqref{eq:kernellog}. The interpolation between grid nodes is performed by a cubic spline. To prevent floating precision errors, we cut the eigenspectrum of the covariance prior $K$ at $10^{-13}$, and of the dataset sample covariance matrix $C$ at $\{10^{-13}, 10^{-12}, 10^{-11}\}$ for $\nu_{max} = \{4, 10, 25\}$. The results for the three datasets are presented in Fig. \ref{fig:zdist} where they are normalized to the input distribution to appreciate the quality of the reconstruction. A more global picture including the extrapolation in $\nu$ space is provided in Figs. \ref{fig:closure_test1} - \ref{fig:closure_test3}. 


Visually, the goodness-of-fit of the data is excellent. It is interesting to notice that quoting a sensible $\chi^2$ value requires additional work that is not needed to construct the posterior. Indeed, $\chi^2 = r^T C^{-1} r$ where $r$ is the vector of residuals. One will notice that in our example, $C$ is in principle non-invertible since -- up to the floating precision regulator -- the data at $\nu = 0$ has zero uncertainty due to the fixed normalization of the input function. Even if one excluded this point from the $\chi^2$ definition, there would still remain eigenvalues on the verge of floating point precision in the covariance of the data that we are fitting in this closure test. 
Furthermore, small eigenvalues of the covariance matrix may be poorly estimated due to the limited number of samples used to construct the dataset. This problem is nonexistent in the posterior distribution construction, since eqs. \eqref{eq:postmean} and \eqref{eq:postcov} do not involve directly $C^{-1}$, but always $[C + BKB^T]^{-1}$ where the prior eliminates the issue.\footnote{To understand the regularization effect of adding $BKB^T$, we can consider the highly degenerate case where the dataset is only made of a single measurement at $\nu = 0$, with 0 uncertainty. The covariance matrix is then $C = (0)$, obviously strictly non-invertible. Using a simple integration rule, if the $x$-grid contains $N$ elements, B is a line vector of $N$ components $(\frac{1}{N}, \frac{1}{N}, ..., \frac{1}{N})$, and $BKB^T = \frac{1}{N^2} \sum_{i,j} K_{i,j}$. This quantity is clearly non-zero since each $K_{i,j}$ is strictly positive. Therefore $C + BKB^T$ is invertible.}
Therefore, while we could simply use an extremely tiny floating precision cutoff to construct the posterior, quoting a sensible goodness-of-fit estimate requires a stronger cutoff on the eigenvalues of $C$. A cutoff of $10^{-6}$ leads to values of the absolute $\chi^2$ significantly smaller than 1. 

When $\nu_{max}$ increases, the extrapolation region shrinks and the reconstruction in $x$ becomes faithful to the input distribution up to increasingly small values of $x$. We quote in Table \ref{tab:tab_xnu} the value of $x$ at which our reconstruction has a standard deviation 50\% larger than the input distribution depending on the value of $\nu_{max}$. We find approximately
\begin{equation}
x \approx 2/\nu_{max}\,,
\end{equation}
which follows the usual known heuristic of the relationship between $\nu_{max}$ and the resolution of the reconstruction in $x$. It is important to remember that this uncertainty is obtained within the procedure that we have detailed before, and tuning the hyperparameters differently -- especially $(\sigma^2, l)$ -- would lead to a different factor in front of $1/\nu_{max}$. We refer to an upcoming further work where a more general study of various kernels and strategies to handle the hyperparameters are explored.

\begin{table}[]
\centering
\bgroup
\def\arraystretch{2.5}
\begin{tabular}{c c}
$\nu_{max}$ & \parbox{5cm}{Value of $x$ such that reconstruction uncertainty is 50\% larger than input uncertainty}  \\
\hline
$\nu_{max} = 4$  & $x = 0.43$  \\
\hline
$\nu_{max} = 10$  & $x = 0.22$  \\
\hline
$\nu_{max} = 25$  & $x = 0.09$ 
\end{tabular}
\caption{Relationship between $\nu_{max}$ and the growth of uncertainty in the $x$ reconstruction.}
\label{tab:tab_xnu}
\egroup
\end{table}

At $x \approx 0$, as we have discussed previously, the posterior becomes equal to the prior. It is quite remarkable that the prior, which is selected dynamically by our request of uncertainty in the extrapolation region, ends up being fairly identical over all cases despite the notable increase in the range in $\nu$ and decrease of the posterior uncertainty for non-zero $x$. Because of the stability of the prior, the uncertainty at $x = 0$ remains globally identical regardless of the value of $\nu_{max}$. This may however be directly linked to the fact that the uncertainty in $\nu$ of the input distribution is itself very slowly varying.

Considering the behavior of the input distribution in $\nu$, it is clear that if we had not forced the uncertainty to increase in the extrapolation region, we could have produced a much more faithful $x$ reconstruction at smaller values of $\nu_{max}$. However, it seems unwarranted to assume that the first principles parton distribution would have the simple features of this two-parameter input distribution. This exercise points out again how large $\nu$ and small $x$ go hand-in-hand. We must therefore exercise caution to avoid bias introduced by using models that infer large $\nu$ behavior from small $\nu$ data.

\section{Application to real data}

\begin{figure*}
\centering
\includegraphics[width=0.8\linewidth]{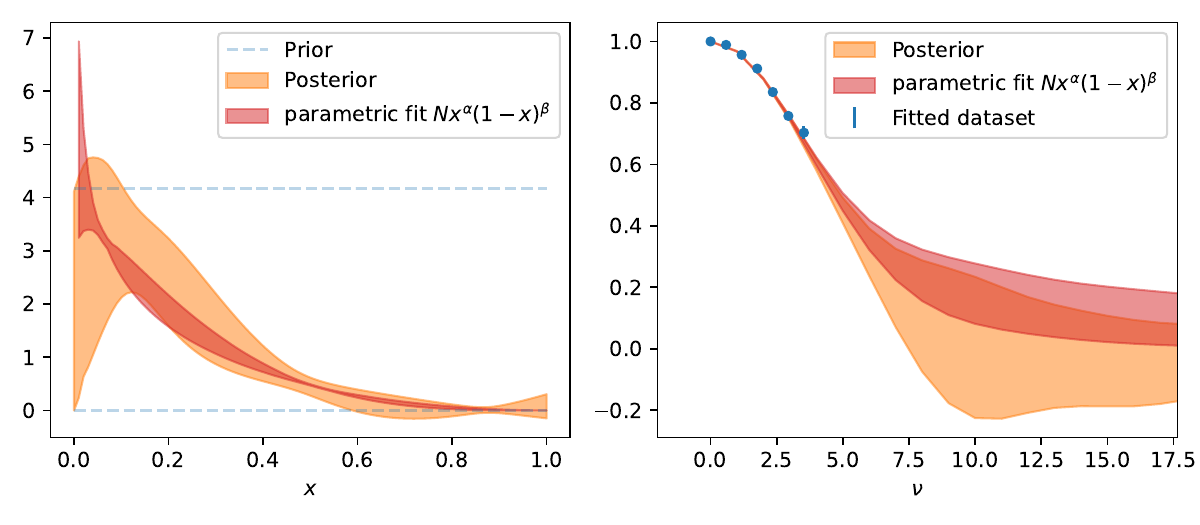}
\caption{Non-parametric (orange) and parametric (red) reconstruction from the real part of the $z = 3a$ unpolarized PDF data published in \cite{Egerer:2021ymv}.}
    \label{fig:real_data_1}
\end{figure*}

\begin{figure*}
\centering
\includegraphics[width=0.8\linewidth]{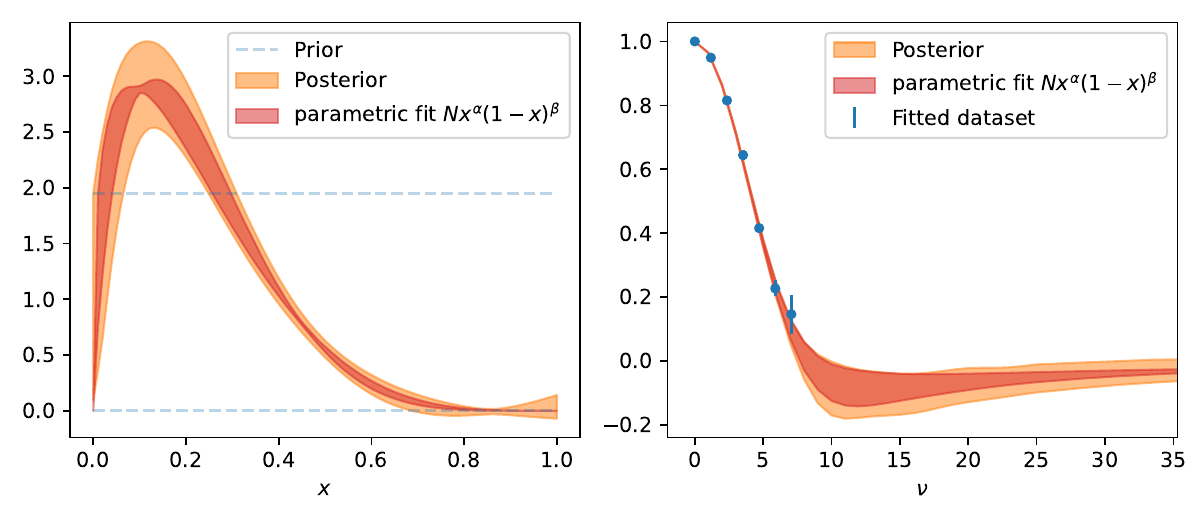}
\caption{Same as Fig. \ref{fig:real_data_1} for the real part of the $z = 6a$ data.}
    \label{fig:real_data_2}
\end{figure*}

\begin{figure*}
\centering
\includegraphics[width=0.8\linewidth]{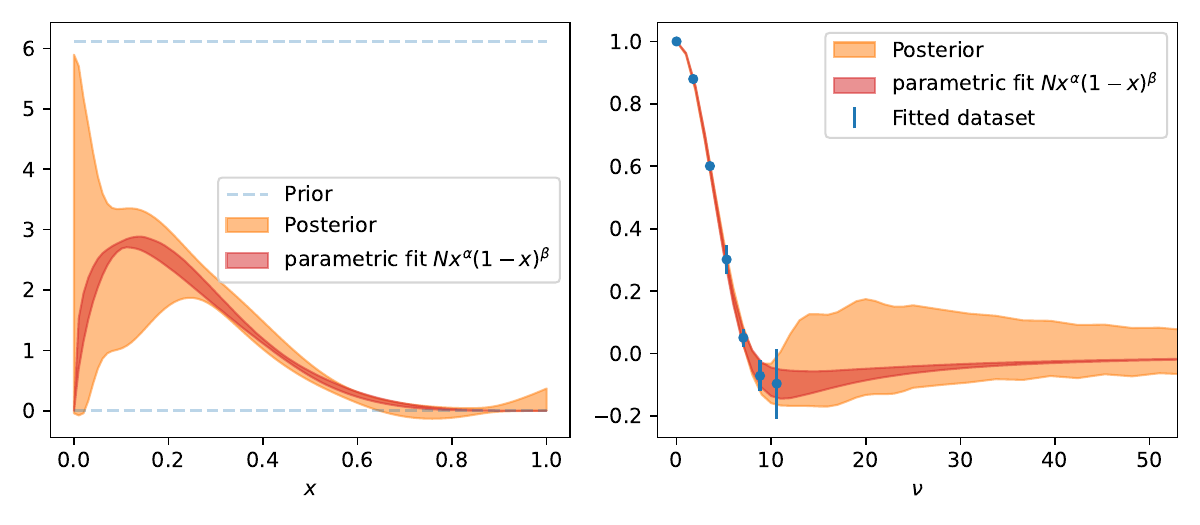}
\caption{Same as Fig. \ref{fig:real_data_1} for the real part of the $z = 9a$ data.}
    \label{fig:real_data_3}
\end{figure*}

\begin{figure*}
\centering
\includegraphics[width=0.8\linewidth]{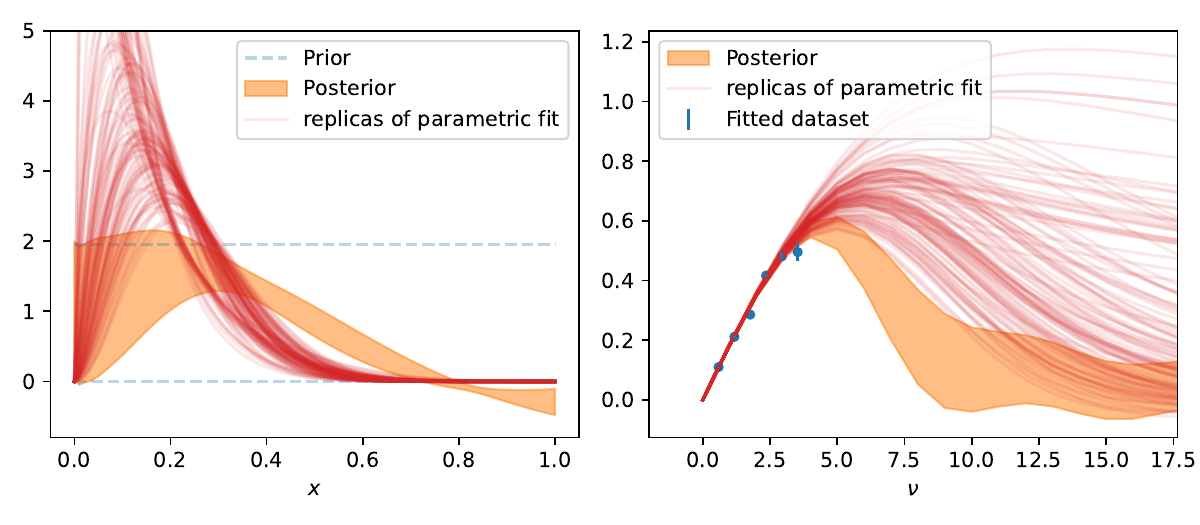}
\caption{Non-parametric (orange) and parametric (red) reconstruction from the imaginary part of the $z = 3a$ unpolarized PDF data published in \cite{Egerer:2021ymv}. The parametric fit is noticeably bimodal in the parameters $(\alpha, \beta)$. For this reason, we prefer to present the parametric fit in terms of replicas.}
    \label{fig:imag_data_1}
\end{figure*}

\begin{figure*}
\centering
\includegraphics[width=0.8\linewidth]{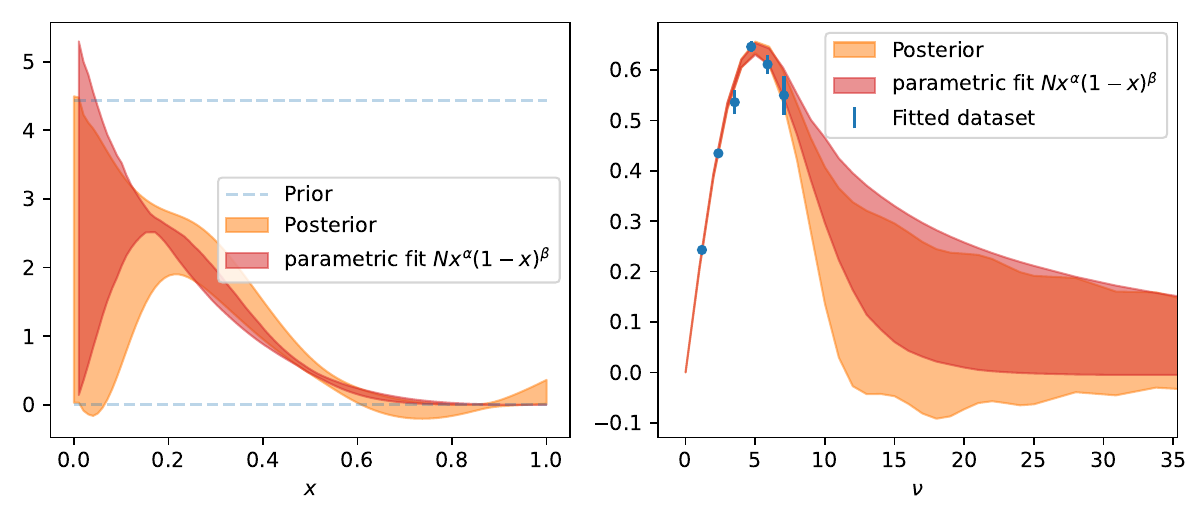}
\caption{Same as Fig. \ref{fig:imag_data_1} for the imaginary part of the $z = 6a$ data. As the parametric fit is distributed approximately normally, we revert to an error band of one standard deviation.}
    \label{fig:imag_data_2}
\end{figure*}

\begin{figure*}
\centering
\includegraphics[width=0.8\linewidth]{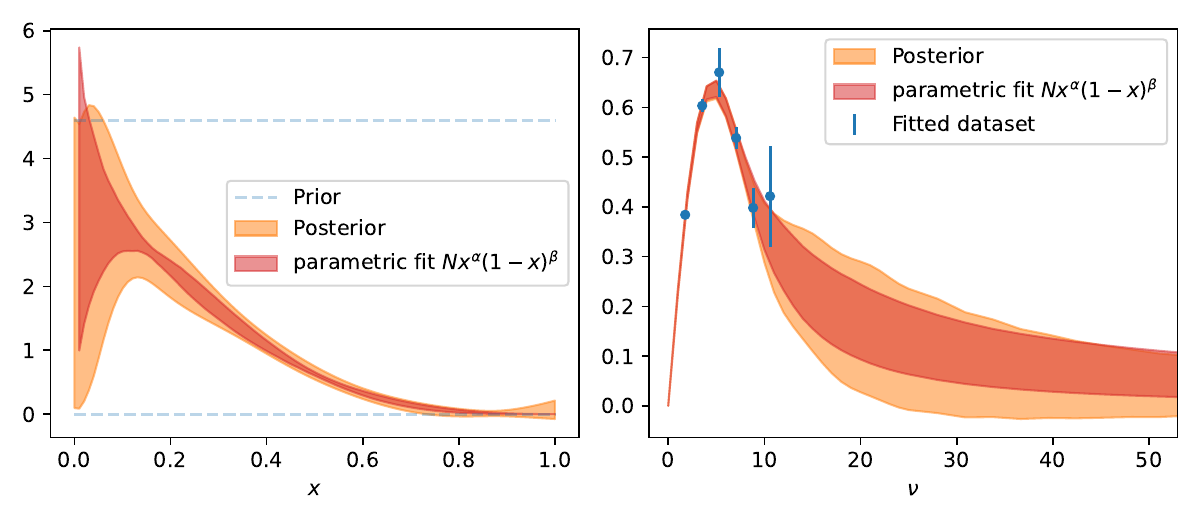}
\caption{Same as Fig. \ref{fig:imag_data_1} for the imaginary part of the $z = 9a$ data.}
    \label{fig:imag_data_3}
\end{figure*}

We apply our procedure to the unpolarized PDF data published in \cite{Egerer:2021ymv} with a lattice spacing of $a = 0.094$ fm, a pion mass of 358 MeV and a maximal proton momentum of $P = 2.47$ GeV. We present results for $z = 3a$ (0.28 fm), $6a$ (0.56 fm), and $9a$ (0.85 fm) for the real parts in Figs. \ref{fig:real_data_1} - \ref{fig:real_data_3}, and the imaginary parts in Figs. \ref{fig:imag_data_1} - \ref{fig:imag_data_3}. To isolate the effect of the Fourier transform uncertainty and not introduce the question of the perturbative matching which comes with its own separate uncertainty, we Fourier transform directly the object $M(\nu, z^2)$ defined in eq. \eqref{eq:matelem}. Therefore, the various reconstructions at different values of $z$ exist at different scales. For larger $z$, this translates into less peaked distributions small $x$, as can be inferred from the effect of the DGLAP evolution equation. As $z$ increases while the largest momentum $P$ is kept fixed, the maximal value of $\nu = P \cdot z$ in the fitted dataset increases likewise.

This time, we apply a floating point precision cutoff on the eigenvalues of the prior covariance $K$ and the sample data covariance $C$ of $10^{-13}$ in all cases.

We compare our reconstruction to the still largely used simple Ansatz:
\begin{equation}
f_q(x) = N x^\alpha (1-x)^\beta\,.
\end{equation}
Because it is a non-linear function of its free parameters, fitting this simple Ansatz is in fact considerably more challenging numerically than the procedure we advocate here. It is both lengthier and suffers from numerical instabilities, as explained in \cite{Bhattacharya:2024qpp}. In the latter, the authors were forced to restrict the range of allowed values of $N$ with a direct noticeable effect on the fit result. In our case, for the real parts, we opt for a stronger cutoff of $10^{-7}$ on the eigenvalues of the data covariance matrix compared to the $10^{-13}$ used in our non-parametric procedure. On the other hand, the smallest eigenvalues in the imaginary part are already larger than $10^{-7}$, so we do not apply any cutoff. Whereas the result of the parametric fits are generally approximately normally distributed, it is noticeably not the case for the imaginary part at $z = 3a$, a situation we will discuss further.


\begin{figure*}
\centering
\includegraphics[width=0.8\linewidth]{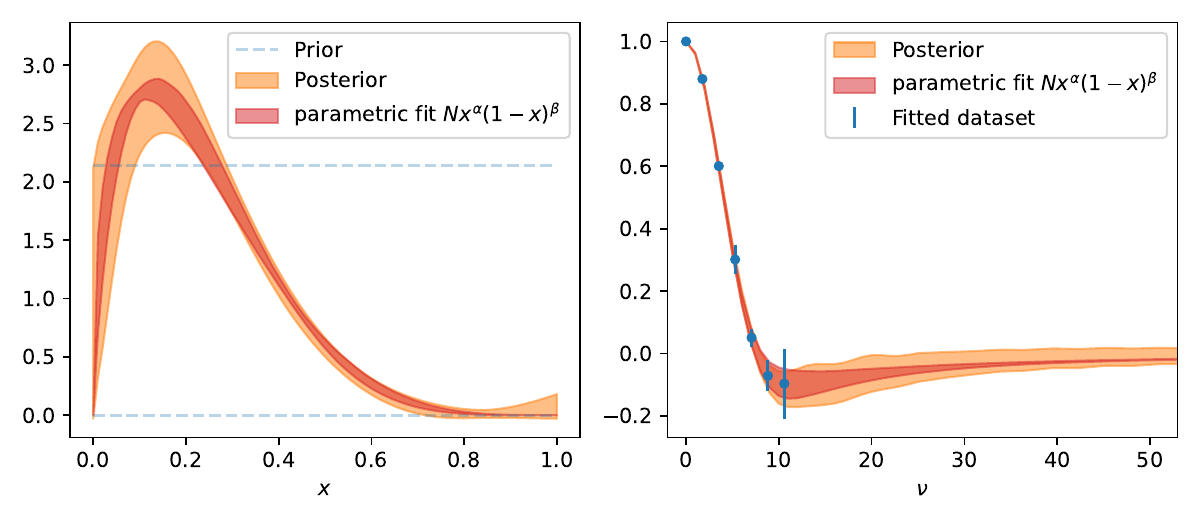}
\caption{Same as Fig. \ref{fig:real_data_3} for the real part of the $z = 9a$ data, but the hyperparameter $\sigma^2$ is fixed according to the former-to-last datapoint.}
    \label{fig:real_data_4}
\end{figure*}

The parametric reconstructions present a well-known issue of strong tightening of the uncertainty at some ``nodes'', especially worrying at small $x$. It is very visible for $x \sim 0.05$ in Fig. \ref{fig:real_data_1} where the parametric uncertainty becomes considerably smaller than the non-parametric one. Whereas it is well understood that in either formalism, the uncertainty for $x \sim 0$ is entirely model or prior-driven, it seems that -- especially when the range in Fourier frequencies is small -- our non-parametric procedure provides a much more valuable uncertainty quantification down to smaller values of $x$ than the parametric form.

The situation for $z = 9a$ in Fig. \ref{fig:real_data_3} is particularly interesting. Whereas the parametric uncertainty is narrow and comparable to the one achieved for $z = 6a$, the non-parametric one has considerably worsened. This may sound paradoxical when the range in $\nu$ has increased. But the uncertainty of the last datapoint for $z = 9a$ is significantly larger than for $z = 6a$. In the parametric fit, due to its large uncertainty which suppresses its contribution to the $\chi^2$, this point matters very little. On the other hand, the $\sigma^2$ hyperparameter of the non-parametric reconstruction is set so that the reconstruction spreads out in the extrapolation region depending on the uncertainty of the last point. Since the uncertainty of the last point is likely excessive, due to a poor lattice signal-to-noise at this large $P$ and $z$ values, we may want to not base our extrapolation on this point. We show in Fig. \ref{fig:real_data_4} the result if the prescription on $\sigma^2$ is applied using the former-to-last datapoint instead. Both the parametric and non-parametric then coincide well. However, the reconstruction of Fig. \ref{fig:real_data_3} is in principle data-driven, as it reproduces the sharp uncertainty increase present in the dataset.

When $x = 1$, in the absence of any constraint requiring that the non-parametric reconstruction goes to 0, the uncertainty tends to increase. Indeed, the information at $x = 1$ is contained in the large Fourier frequencies that are less constrained by the data. Let us note that the mock dataset of the previous section did in fact constrain surprisingly well the large $x$ region. As we have mentioned previously, one can easily enforce this additional condition if desired.

The case of the imaginary part for $z = 3a$ is also noteworthy in Fig. \ref{fig:imag_data_1}. The parametric reconstruction is extremely unstable: several non-linear minimizers regularly failed to converge due to the extremely bad conditioning of the fitting. It is the only case in this study where the distribution of best-fit parameters we obtain by sampling replicas of the dataset deviates strongly from a normal distribution. It is noticeably bimodal around the modes $(\alpha = 1.5, \beta = 9)$ and $(\alpha = -0.5, \beta = 4)$. The latter clearly corresponds to a better phenomenological picture but is less frequent in the replicas. The two modes of the distribution can be distinguished in the replicas presented on Fig. \ref{fig:imag_data_1} by a branching effect for $x \sim 0.3$ and a different trend at large $\nu$. Overall, the parametric fit is simply extremely unreliable from a numerical point of view on such a small range in $\nu$, where the Fourier data mostly presents a simple monotonical growth. The non-parametric reconstruction does not suffer from any of those issues. There again, it is easy to tune the value of $\sigma^2$ if the uncertainty of the reconstruction in $\nu$ is deemed not quite large enough for instance. In the imaginary part at $z = 6a$ and $9a$, we find good compatibility between both methods, with once again an unphysical tightening of the parametric uncertainty at small $x$. Here as well, it seems that our non-parametric reconstruction provides much more meaningful uncertainty quantification down to much smaller values of $x$ than the parametric one, and also avoids serious numerical issues as the one in Fig. \ref{fig:imag_data_1}.


\section{Applicability of this procedure to other inverse problems}\label{sec:other_inverse}

Although we have focused so far on the $x$-dependence of one-dimensional parton distributions from the short-distance factorization of non-local matrix elements computed on the lattice, our procedure can be applied in other contexts. A first obvious application is the computation of parton distribution functions from the same matrix elements in the LaMET formalism \cite{Ji:2013dva, Ji:2014gla}. Constructing the quasi-PDF also requires a reconstruction of $x$-dependent objects from limited Fourier information. It has become customary in many recent works to complete the tail of missing Fourier modes by a simple parametric model \cite{Ji:2020brr}, inspired by general expectations on the behavior of the non-local matrix element for large separations in $z$. A popular simple model with 3 free parameters is sometimes applied as early as $\nu \approx 5$ \textit{e.g.} \cite{Gao:2022uhg,Ding:2024hkz}. The non-parametric procedure advocated here could help reduce the model-dependence introduced in such calculations. Since quasi-PDFs are known to have formally support in the interval $[-\infty, \infty]$ for finite hadron momentum, one needs to extend the support beyond $x \in [0,1]$. However, at reasonably large momentum, a much smaller support is likely sufficient. 

\begin{figure}
\centering
\includegraphics[width=0.99\linewidth]{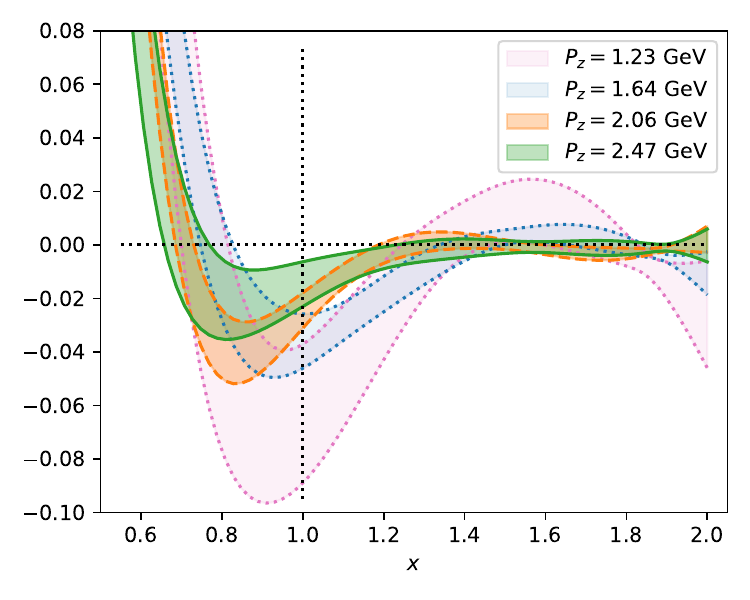}
\caption{Non-parametric reconstruction of the quasi-PDF $q(x, P_z)$ from the real part of unpolarized PDF data published in \cite{Egerer:2021ymv}. The quasi-PDF exhibits only tiny contributions in the region $[1,2]$, which are reduced when $P_z$ increases.}
    \label{fig:quasiPDF}
\end{figure}

We reanalyze the unpolarized PDF data of \cite{Egerer:2021ymv} used in the previous section in the quasi-PDF formalism, namely performing the inverse Fourier transform at fixed hadron momentum $P_z$ instead of fixed non-local separation $z$. We apply exactly the procedure advertised before, with the only modification that we allow the support in $x$ to extend to $[0,2]$. We present the results for four momenta $P_z = 1.23$~GeV up to 2.47 GeV in Fig. \ref{fig:quasiPDF}. For all three largest momenta, our data in Ioffe time extends roughly up to the same $\nu_{max} \approx 9$. For the smallest momentum, it only extends up to $\nu_{max} \approx 6$ which explains in part the larger uncertainty. We are mostly interested to see how the reconstruction behaves outside of the range $[0,1]$. As observed in Fig. \ref{fig:quasiPDF}, the contribution outside of $[0,1]$ is very small in magnitude (the quasi-PDF reaches a peak of the order of 3 when $x \sim 0.15$). Furthermore, we retrieve the expected feature that it decreases when $P_z$ increases.

Another extension of the proposed procedure concerns multi-dimensional parton distributions. For instance, the short-distance factorization of non-forward non-local matrix elements allows to reconstruct the three-dimensional generalized parton distributions (GPDs) \cite{Muller:1994ses, Ji:1996nm, Radyushkin:1997ki}. Sparing the reader the introduction of lengthy new notations, the problem presents itself in the double distribution representation \cite{Radyushkin:1998es, Radyushkin:2023ref} under the form of a two-dimensional Fourier transform with an additional external kinematic variable:
\begin{equation}
M(\nu, \bar{\nu}, t) = \int_\Omega \mathrm{d}\alpha \mathrm{d}\beta\,e^{i(\beta \nu + \alpha \bar{\nu})} f_q(\beta, \alpha, t)\,, \label{eq:2DFT}
\end{equation}
where the 2D support $\Omega$ is defined by $\{(\alpha, \beta)\,|\,|\alpha|+|\beta| < 1\}$. The problem of reconstructing $f_q(\beta, \alpha, t)$ shares a lot of similarities with the one we have addressed before. The increase in dimension in not really an issue \textit{per se} as eq. \eqref{eq:2DFT} can still be written as a simple linear transform similar to the one we have dealt with so far. However, more thought needs to be devoted to the physics-driven choice of prior kernels and hyperparameters. In the one-dimensional case, we know from our understanding of partonic physics 
that the small-x and large-x regions are uncorrelated and thus we have chosen a kernel that has these properties.
This feature translates directly into the $\beta$-space. 
But in terms of the two-dimensional $(\beta, \alpha)$ plane, should we allow more complicated physically motivated correlations? For instance, rather than decoupling the large $\beta$ from small $\beta$, could it be more appropriate to decouple $(\beta \approx 0, \alpha \approx 0)$ from the rest of the kinematic domain? To imprint an appropriate behavior in the extrapolation in $t$ also requires tests that go beyond this short article, and which are of general interest for the phenomenology of GPDs.

Another important source of first-principle information on parton distributions is obtained from Mellin moments:
\begin{equation}
\langle x^{s-1}\rangle_q = \int_{-1}^1 \mathrm{d}x\,x^{s-1}f_q(x)\,.
\end{equation}
On the lattice, whether using local or non-local operators, a variable number of integer values of $s \in \{1, 2, 3, ...\}$ are usually accessed. Although this problem shares deep connections with the one we have addressed before (Mellin moments are derivatives of the Fourier transform at $\nu = 0$), the extrapolation strategy must be revised. If one considers the Mellin space with continuous values of the variable $s$, then the bulk of the uncertainty at small $x$ is now contained in the extrapolation region at small $s < 1$. The physics-driven prescription we have developed in this paper is not directly applicable in this context. Further examination of this important channel for lattice calculations is necessary.

Finally, the main source of information on parton distributions remains in many cases the QCD factorization of some experimental processes. The problem presents itself differently in that case, because -- to some approximation -- the measured data exists in the same space as the desired construction. Measurement of deeply inelastic structure functions gives access to the following integrals of the function to reconstruct:
\begin{equation}
F(\xi) = \int_{-1}^1 \frac{dx}{|\xi|}\,C\left(\frac{x}{\xi}\right) f_q(x)\,.
\end{equation}
To leading order in perturbation theory, $C(\alpha) = \delta(1-\alpha)$, meaning that the measured data approximately determines the $x$-dependence of the function to reconstruct. At higher order in perturbation theory, the strict equivalence between $\xi$ and $x$ ceases to be valid. However, in the range of $\xi$ covered by experimental data, properties such as the correlation length of the prior can be informed by the data themselves more directly than what Fourier (or Mellin) modes allow to do. We believe sampling over hyperparameters, as studied in \cite{Candido:2024hjt} with synthetic structure function data, is probably the most meaningful strategy.

\section{Conclusion}

We have presented a simple non-parametric method with a physical prescription to reconstruct parton distributions from limited Fourier modes. The effect of our physical prescription on the reconstruction uncertainty is easily understood and tunable if the results are unsatisfactory. We have found that our procedure gives satisfactory results, whether with simulated or real data. Especially when the range in Fourier frequencies is reduced, we have found our uncertainty to be much more in tune with physical expectations than a popular parametric model, which produces artifacts and frequently unreasonably optimistic uncertainties at small $x$.  Our method is also very efficient numerically, and its results can be communicated easily and exactly since the reconstruction is a multivariate normal distribution. We have discussed the relevant features of other related inverse problems in hadronic structure and paved the way for a more systematic study of those cases.

\section*{Acknowledgements}

We thank Yamil Cahuana Medrano and Eloy Romero for stimulating discussions. SZ acknowledges fruitful discussions with Jan Pawlowski, Jonas Turnwald, and Julian Urban. HD was supported in part by U.S. DOE Grant \#DE-FG02-04ER41302 and under the Laboratory Directed Research and Development Program (LDRD 2412) at the Thomas Jefferson National Accelerator Facility for the U.S. Department of Energy. KO was supported in part by U.S.~DOE Grant \mbox{\#DE-FG02-04ER41302}. This project was supported by the U.S.~Department of Energy, Office of Science, Contract \#DE-AC05-06OR23177, under which Jefferson Science Associates, LLC operates Jefferson Lab. This work has benefited from the collaboration enabled by the Quark-Gluon Tomography (QGT) Topical Collaboration, U.S.~DOE Award \mbox{\#DE-SC0023646}. Computations for this work were carried out in part on facilities of the USQCD Collaboration, which are funded by the Office of Science of the U.S.~Department of Energy. The research of SZ was funded, in part, by l’Agence Nationale de la Recherche (ANR), project ANR-23-CE31-0019. For the purpose of open access, the author has applied a CC-BY public copyright license to any Author Accepted Manuscript (AAM) version arising from this submission. This work was performed in part using computing facilities at William \& Mary which were provided by contributions from the National Science Foundation (MRI grant PHY-1626177), and the Commonwealth of Virginia Equipment Trust Fund. In addition, this work used resources at NERSC, a DOE Office of Science User Facility supported by the Office of Science of the U.S. Department of Energy under Contract \#DE-AC02-05CH11231, as well as resources of the Oak Ridge Leadership Computing Facility at the Oak Ridge National Laboratory, which is supported by the Office of Science of the U.S. Department of Energy under Contract No. \mbox{\#DE-AC05-00OR22725}. 
The authors acknowledge support as well as computing and storage resources by GENCI on Adastra (CINES), Jean-Zay (IDRIS) under project (2020-2024)-A0080511504.
The software codes {\tt Chroma} \cite{Edwards:2004sx}, {\tt QUDA} \cite{Clark:2009wm, Babich:2010mu}, {\tt QPhiX} \cite{QPhiX2}, and {\tt Redstar} \cite{Chen:2023zyy} were used in our work. The authors acknowledge support from the U.S. Department of Energy, Office of Science, Office of Advanced Scientific Computing Research and Office of Nuclear Physics, Scientific Discovery through Advanced Computing (SciDAC) program, and of the U.S. Department of Energy Exascale Computing Project (ECP). The authors also acknowledge the Texas Advanced Computing Center (TACC) at The University of Texas at Austin for providing HPC resources, like Frontera computing system~\cite{frontera} that has contributed to the research results reported within this paper. The authors acknowledge William \& Mary Research Computing for providing computational resources and/or technical support that have contributed to the results reported within this paper.

\bibliography{apssamp}

\end{document}